\documentclass[%
preprint,
superscriptaddress,
amsmath, 
amssymb,
aps,
floatfix,
]{revtex4-1}
\usepackage{setspace}
\usepackage{lipsum, babel}
\usepackage{printlen}
\usepackage{float}
\usepackage[toc,page]{appendix}
\usepackage{graphicx}
\usepackage{dcolumn}
\usepackage{bm}
\usepackage[colorlinks=true,citecolor=blue,linkcolor=magenta,hypertexnames=false]{hyperref}
\usepackage[utf8]{inputenc}
\usepackage{url}
\usepackage[normalem]{ulem}
\usepackage{upgreek}
\usepackage{siunitx}
\usepackage[mathlines]{lineno}
\usepackage{amsmath}
\usepackage{xr}
\usepackage{verbatim}

\newcommand{\cmt}[1]{{}}
\newif\ifproofread
\newcommand{\changemarker}[1]{%
\ifproofread
\textcolor{red}{#1}%
\else
#1%
\fi
}
\usepackage[usenames,dvipsnames]{color}

\usepackage{xr}
\makeatletter
\newcommand*{\addFileDependency}[1]{
  \typeout{(#1)}
  \@addtofilelist{#1}
  \IfFileExists{#1}{}{\typeout{No file #1.}}
}
\makeatother

\renewcommand\thefigure{{\fontseries{bx}\selectfont \arabic{figure}}} 

\usepackage{multirow}
\usepackage{soul}
\usepackage{layouts}

\begin{document}

\title{$\upmu$-MOPA Architecture for Photonic Integrated Solid State Laser}
\author{Yu Guo}
\affiliation{Department of Electrical and Computer Engineering, Yale University, New Haven, CT 06511, USA}
\affiliation{These authors contributed equally to this work}
\author{Yubo Wang}
\affiliation{Department of Electrical and Computer Engineering, Yale University, New Haven, CT 06511, USA}
\affiliation{These authors contributed equally to this work}
\author{Haoqi Zhao}
\affiliation{Department of Electrical and Computer Engineering, Yale University, New Haven, CT 06511, USA}
\author{Fengyan Yang}
\affiliation{Department of Electrical and Computer Engineering, Yale University, New Haven, CT 06511, USA}
\author{Guangcanlan Yang}
\affiliation{Department of Electrical and Computer Engineering, Yale University, New Haven, CT 06511, USA}
\author{Hao Xie}
\affiliation{Department of Electrical and Computer Engineering, Yale University, New Haven, CT 06511, USA}

\author{Hong X. Tang}
\affiliation{Department of Electrical and Computer Engineering, Yale University, New Haven, CT 06511, USA}
\affiliation{Corresponding author: hong.tang@yale.edu}
\date{\today}
\begin{abstract}
\vspace{12pt}
Diode-pumped solid-state (DPSS) lasers play a central role in modern photonics owing to their exceptional efficiency and ability to extend spectral coverage beyond the reach of semiconductor diodes. These attributes have enabled breakthroughs in precision metrology, quantum optics, and coherent communications. However, bringing the proven advantages of DPSS gain media such as Nd:YAG onto an integrated photonic platform has remained difficult, largely due to inefficient pump utilization and limited power-scaling in chip-scale implementations. Here, we demonstrate the first photonic-integrated Nd:YAG laser–amplifier system that overcomes these challenges with a micro-chip based master-oscillator–power-amplifier ($\upmu$-MOPA) architecture. The seed laser, employing a double-resonant microring resonator, could reach a threshold as low as 2.9 $\upmu$W. The single-pass waveguide amplifier, when optimized separately, provides up to 46.6 dB small-signal gain. Combining the low-threshold seed with cascaded waveguide amplifiers, the integrated $\upmu$-MOPA delivers more than 12 dBm of amplified continuous-wave output power. These results establish Nd:YAG waveguide integration as a practical route to compact and high-performance solid-state light sources. 

\end{abstract}
\maketitle

Diode-pumped solid-state (DPSS) lasers~\cite{fan2002diode,byer1988diode, koechner2013solid} are essential to modern photonics, offering high pump efficiency, narrow linewidth, diffraction-limited beams, and access to key spectral regions beyond the reach of semiconductor diodes. These traits have enabled advances in precision metrology~\cite{Hakobyan2017OFC,Gurel2017OFC}, quantum control, coherent communications, spectroscopy~\cite{mayer2017watt,fritsch2022dual,Phillips2023dual}, medicine~\cite{DESANTIS2018dental,Baoyi2024LADD}, and materials processing~\cite{liu2021review}. Among DPSS media, Nd:YAG stands out: its four-level energy structure~\cite{geusic1964laser} supports quantum-defect-limited power conversion efficiency of \SI{76}{\percent} under \SI{808}{\nano\meter} pumping while providing long-term stability~\cite{Tidwell1991eff60,Qi2005eff62,Zhang2019eff60}. As a result, DPSS Nd:YAG sources remain a gold standard against which other solid-state lasers are compared.

The growing push toward integrated photonic systems now raises a natural question: can the virtues of DPSS lasers be brought onto a chip? Conventional DPSS architectures~\cite{huber2010solid} rely on bulk cavities, discrete crystals, and high-power pumping,  making direct translation into integrated platforms non-trivial. Yet achieving this miniaturization would establish a new class of compact, efficient, and stable on-chip lasers for next-generation photonic systems~\cite{wang2023photonic,yang2024titanium,Yin2021ErLNOI,Liu2024HybridErbiumLaser,Luo2022YbLNMLaser,singh2025sub2w}. Heterogeneous photonic integration provides a powerful route forward by combining crystalline gain media with ultra-low-loss waveguides on a common substrate~\cite{lu2024emerging}. Integrated Ti:sapphire lasers have validated this strategy in the visible regime with low pump thresholds~\cite{wang2023photonic, yang2024titanium, wang2024wafer}, while Er-doped waveguide amplifiers extend gain into the telecom band~\cite{liu2022photonic, bradley2011erbium}. In contrast to Ti:sapphire and erbium-based gain medium, Nd:YAG offers significantly higher gain efficiency, making it a strong candidate for integrated platforms for achieving both low lasing threshold and high gain amplification~\cite{veisz2025waveform,Deng2018higheff}. Moreover, Nd:YAG emits near 1 µm, which bridges the spectral gap between visible and telecom bands and enables key applications such as optical trapping and ion control~\cite{niffenegger2020integrated, taylor2013biological}, as well as biomedical~\cite{menazea2020precipitation,Kranz2023OCT,Shi2025Rosacea}, metal processing~\cite{gurusami2020strengthening, leone2018heat}, and eye-safe sensing applications. Recent attempts to miniaturize Nd:YAG lasers using micro-transferred thin films~\cite{li2024heterogeneous, li2023optically} have demonstrated on-chip gain, but their power performance has been constrained by limited pump utilization under a resonantly-pumped ring laser geometry.

Here, we demonstrate the first hybrid diode-pumped Nd:YAG laser suite on a silicon-nitride platform that enables a micro-chip based master-oscillator–power-amplifier ($\upmu$-MOPA) laser architecture with both low pump threshold and scalable output power. Single-crystal Nd:YAG is directly bonded onto ultra-low-loss Si$_3$N$_4$ waveguides engineered to co-confine 808 nm pump light and the 1.064 µm signal. A lensed fiber array enables efficient coupling from commercial pump diodes, preserving the canonical DPSS architecture in a small footprint. The ``master'' laser of our on-chip MOPA is a single-ring double-resonant laser designed for single-mode lasing, capable of reaching a low threshold of 2.9 $\upmu$W. 
This wavelength defining ``master'' laser is then amplified by a follow-on single-pass waveguide ``amplifier'', whose output can be further boosted by cascading additional amplification stages. The complete system yields an integrated MOPA architecture that surpasses integrated Ti:Sapphire and Er-doped devices in photon-conversion efficiency and power density, thus setting a new benchmark for chip-scale solid-state lasers.

\vspace{0.6cm}
\noindent\textbf{$\upmu$-MOPA: Modeling and Conversion Efficiency}

\begin{figure*}[t]
\centering
\includegraphics[trim={0cm 0cm 0cm 0cm},clip,width=1\linewidth]{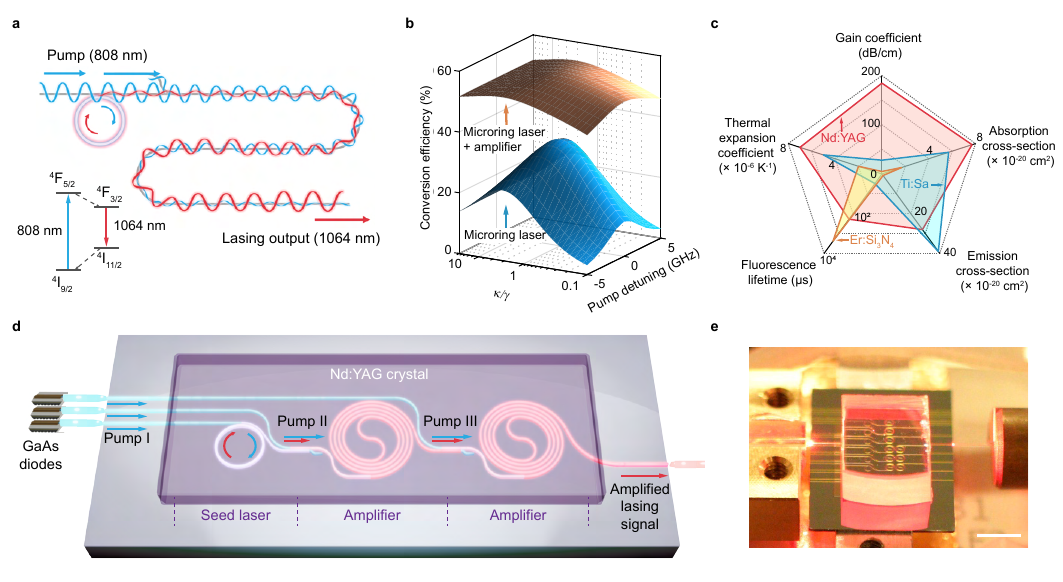}
\caption{{\fontseries{bx}\selectfont Chip-integrated Nd:YAG $\upmu$-MOPA lasers with high photon conversion efficiency.}
{\fontseries{bx}\selectfont a}, Conceptual schematic of the integrated $\upmu$-MOPA laser. The single Si$_3$N$_4$ microring is modeled as two resonant cavities simultaneously aligned with the pump and lasing wavelengths. The ring resonator generates a low-threshold seed laser that is subsequently amplified in spiral waveguides. Inset: Simplified Nd:YAG energy-level diagram showing optical pumping at 808 nm and stimulated emission at 1064 nm.
{\fontseries{bx}\selectfont b}, Conversion efficiency (\textit{P}$_\text{lasing}$/\textit{P}$_\text{pump}$) dependence on pump detuning and coupling condition parameter ($\kappa$/$\gamma$, the ratio of coupling rate to intrinsic loss rate). The lower surface corresponds to a microring laser pumped with 10 mW on-chip power, while the upper surface represents a $\upmu$-MOPA configuration with distributed pumping (1 mW to the microring laser and 9 mW to the amplifier).
{\fontseries{bx}\selectfont c}, Comparative radar chart of gain-related parameters among representative solid-state laser materials. The doping concentrations used for gain evaluation are $1.52\times10^{20}\ \mathrm{cm^{-3}}$ for Nd:YAG, $1.65\times10^{19}\ \mathrm{cm^{-3}}$ for Ti:Sa, and $3.25\times10^{20}\ \mathrm{cm^{-3}}$ for Er:Si$_3$N$_4$.
{\fontseries{bx}\selectfont d}, Schematic of the integrated Nd:YAG $\upmu$-MOPA architecture, showing cascaded lasing and dual-stage signal amplification driven by GaAs diode pumps.
{\fontseries{bx}\selectfont e}, Photograph of a fabricated device coupled with lensed fiber arrays. Scale bar: 3 mm.}
\label{fig1}
\end{figure*}

While both Fabry–Pérot and microring resonators have been explored for on-chip solid-state lasers, microring geometries are particularly attractive for their compact footprint and high cavity finesse. Nevertheless, for DPSS lasers, achieving lasing efficiency approaching the quantum-defect limit with a single ring resonator is challenging. Efficient pumping of the ring requires critical coupling of the pump, with the external coupling rate matched to the cavity internal loss ~\cite{bogaerts2012silicon}, as shown by the lower contour in Fig.~\ref{fig1}b. Simultaneously, the lasing mode should ideally be over-coupled to obtain high extraction efficiency. These impose stringent design spaces that are difficult to realize. The challenge is further compounded under high-power operation, where commercial multimode pump diodes possess broad spectral bandwidths and free-spectral-range mismatches with the microring resonance, preventing efficient energy transfer and leaving much of the pump power uncoupled from the cavity. 
The $\upmu$-MOPA architecture overcomes these limitations by cascading a single ring resonator with an on-chip waveguide amplifier, with the pump power distributed between the two elements, as shown in Fig.~\ref{fig1}a. In this configuration, the ring resonator provides a low-threshold seed laser that is subsequently coupled into the amplifier. A sufficiently long, low-loss waveguide enables the amplifier to efficiently boost the seed power, with a conversion efficiency that can, in principle, approach the quantum-defect limit~\cite{koechner2013solid}. Consequently, high lasing efficiency is maintained across a broad design parameter space, and is more insensitive to coupling strength or pump detuning, as shown by the upper contour in Fig.~\ref{fig1}b. Moreover, the intrinsic scalability of our heterogeneous integration platform provides a direct route to cascaded amplification stages, offering a straightforward pathway to higher output powers.

\vspace{.6cm}

The high efficiency of our integrated Nd:YAG platform arises from the combination of high material gain and low-loss photonic integration. Figure~\ref{fig1}a (inset) presents the simplified energy-level diagram of Nd:YAG. Its four-level energy state, in which the lower laser level lies above the ground state and rapidly depopulates via nonradiative decay, suppresses reabsorption losses compared with quasi-three-level media such as Er-doped lasers~\cite{Liu2024HybridErbiumLaser,bradley2011erbium}. Combined with its large absorption and emission cross-sections~\cite{Liu2005PhotonicDevices}, Nd:YAG \changemarker{exhibits a maximum} small-signal gain exceeding $180\,\mathrm{dB\,cm^{-1}}$ \changemarker{under saturation pumping} (Fig.~\ref{fig1}c), standing at the forefront of high-gain laser crystals. With Si$_3$N$_4$ propagation loss below \SI{0.3}{\decibel\per\centi\meter}, the large gain-to-loss ratio yields efficient pump-to-signal conversion and high slope efficiency.

The conceptual layout of the integrated Nd:YAG $\upmu$-MOPA is illustrated in Fig.~\ref{fig1}d. The active Nd:YAG gain layer is directly bonded on top of the passive photonic layer (Methods 1)
. The passive stack comprises a silicon substrate, a 4\,µm-thick oxide bottom cladding, and 380\,nm-thick Si$_3$N$_4$ ultralow-loss waveguides. A pulley-coupled microring resonator is employed as the master oscillator, with its coupling condition carefully engineered to simultaneously provide efficient pump (Pump I) coupling and high signal extraction. The seed laser is then extracted through a directional coupler, combined with the subsequent pump stages (Pump II and III), and routed into cascaded spiral waveguides for power amplification. All three pump beams are provided by commercial laser diodes and coupled into the chip simultaneously via a lensed fiber array. To maintain stable operation of the master oscillator, the device is mounted on a thermoelectric cooler. A photograph of a fully fabricated device is shown in Fig.~\ref{fig1}e.

\vspace{0.6cm}
\noindent\textbf{Integrated high gain Nd:YAG amplifier} 

Optical amplifiers enable high output power while preserving the coherence of the seed laser. In rare-earth–doped media, amplification originates from stimulated emission between discrete, long-lived atomic energy levels, which yields  intrinsically low noise compared to broadband semiconductor gain media~\cite{liu2022photonic, Osornio-Martinez:25}. To characterize the amplification performance of our integrated Nd:YAG platform, we fabricated standalone spiral waveguides and characterized their response using the setup shown in Extended Data Fig.~1a. The amplifier employs a \SI{2}{\micro\meter}-wide waveguide designed for strong pump confinement and a pump-to-signal mode overlap factor exceeding 92\%, ensuring efficient energy transfer under low pump power (Supplementary Information). A commercial laser diode centered at \SI{1.064}{\micro\meter} provides the signal input, which is combined with the pump via separate input channels of a fiber-based wavelength-division multiplexer (WDM) and coupled into the device in a co-propagating configuration. The measured facet losses are 6.8 ± 0.3 dB for the pump and 5.7 ± 0.2 dB for the signal, \changemarker{mainly due to strong mode confinement in the thin, air-clad waveguide, and can be reduced in future implementations using improved spot-size converters~\cite{Liang2022SiNEdgeCoupler,Brunetti2023SiNSpotSize}}. 

\begin{figure*}[t]
\centering
\includegraphics[width=\linewidth]{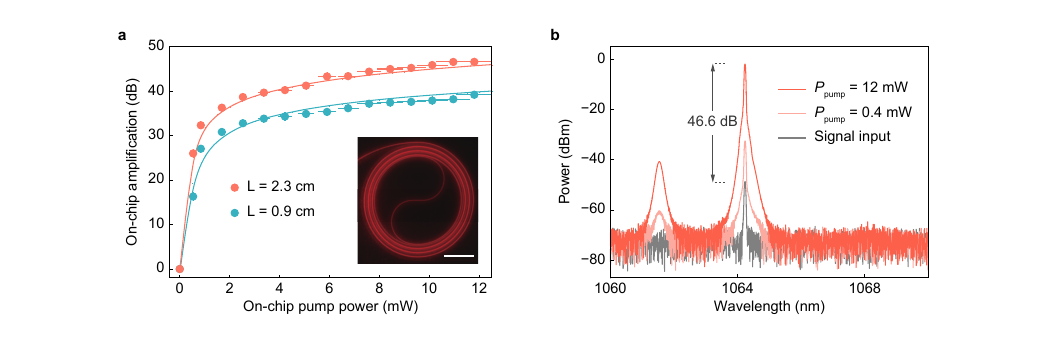}
\caption{{\fontseries{bx}\selectfont Integrated Nd:YAG  waveguide optical amplifiers.}
{\fontseries{bx}\selectfont a}, Measured amplification as a function of pump power for two waveguide lengths. Solid lines: numerical model; points: measurements. Inset: optical microscope image of a \SI{2.3}{\centi\meter}-long spiral amplifier during amplification measurement. Scale bar, \SI{200}{\micro\meter}.
{\fontseries{bx}\selectfont b}, Output spectra of the amplified signal under increasing pump powers, showing a net gain of 46.6 dB and an amplified spontaneous emission suppression exceeding 40 dB. 
}
\label{fig2}
\end{figure*}

To assess the achievable gain under near-complete pump absorption, we fabricated spiral waveguides of varying lengths (Fig.~\ref{fig2}a). Using a 2.3-cm-long spiral waveguide and a small on-chip signal power of \SI{0.02}{\micro\watt}, we observe a net gain of 46.6 dB with only \SI{12}{\milli\watt} of pump power, representing, to the best of our knowledge, the highest reported amplification in an integrated single-stage amplifier based on both rare-earth and transition-metal doped solid state gain media~\cite{liu2022photonic, bradley2011erbium,Bao2024ErDopedTFLNAmp,Wei2025ErbiumLNOIAmp,Zhang2023YbLNOIAmp,Zhang2023ErYbTFLNAmp,Singh2025WattAmplifier}. 
Figure~\ref{fig2}b shows the evolution of the output spectrum as the on-chip pump power is increased. An amplified spontaneous emission (ASE) suppression exceeding 40 dB is observed, indicating highly selective single-wavelength amplification \changemarker{with no evidence of parasitic lasing (Supplementary Fig. S3)}. This level of suppression preserves the spectral purity and coherence of the amplified signal, enabling low noise, narrow-linewidth lasing. 
\changemarker{Under large-signal operation, the device delivers 9.0 dB gain from a \SI{1}{\milli\watt} input, corresponding to a high pump-to-signal conversion efficiency of 52.2\%, as shown in Extended Data Fig.~1b.}

\vspace{0.6cm}
\noindent\textbf{Integrated Nd:YAG microring lasers}

To fully leverage this capability in a self-contained laser system, we next investigate the implementation of the seed laser on-chip. The Nd:YAG seed laser uses a dual-resonance micro-ring resonator that supports cavity modes at both the pump and signal transitions of the Nd:YAG four-level system. This design boosts pump efficiency 
and ensures strong spatial and spectral overlap between the 808 nm pump and 1064 nm signal. Simulations confirm a near-unity pump-to-signal mode overlap, with a tightly confined optical mode volume of less than 1 $\upmu$m$^2$~\cite{wang2024heterogeneous}. 
The micro-ring laser output performance is primarily influenced by the design of the optical couplers. For the pump mode, a critically coupled regime maximizes energy transfer and ensures near-complete absorption within the gain medium. For the signal mode, the optimal coupling condition depends on the targeted laser characteristics: overcoupling yields efficient light extraction and high output power, whereas undercoupling enhances intracavity buildup and minimizes the lasing threshold. To demonstrate both regimes experimentally, we designed devices following these two coupling principles.

\begin{figure*}[t]
\centering
\includegraphics[width=\linewidth]{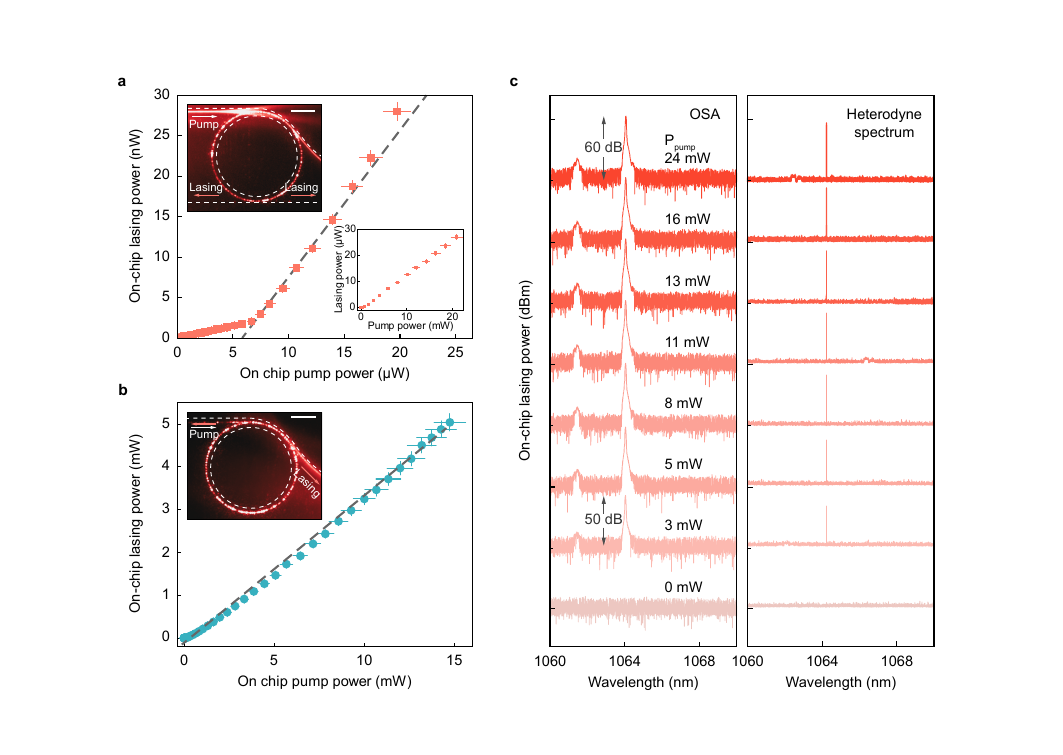}
\caption{{\fontseries{bx}\selectfont High‑efficiency Nd:YAG microring laser.}
{\fontseries{bx}\selectfont a}, Light–light (L–L) curve of a low-threshold device. Upper inset: optical microscope image of the double-resonant microring. Lower inset: L-L curve at higher pump powers.   
{\fontseries{bx}\selectfont b}, L–L curve of a single-port device, showing higher slope efficiency and output power exceeding \SI{5}{\milli\watt}.  
{\fontseries{bx}\selectfont c}, Evolution of single-mode lasing (traces vertically offset by 60 dB per division). Left: optical spectrum analyzer (OSA) trace shows a broadened emission profile limited by the instrument’s resolution bandwidth, whereas right: heterodyne beatnote confirms stable single-mode emission. Scale bar, \SI{150}{\micro\meter}.
}
\label{fig3}
\end{figure*}

In the low-threshold configuration, a two-port coupling scheme is employed, where a phase-matched pulley coupler efficiently injects the pump and a point coupler with a large gap extracts the lasing mode while maintaining a high-Q cavity (Fig.~\ref{fig3}a, inset). To evaluate the intrinsic photonic properties of the integrated Nd:YAG platform, we independently measured the optical losses at the pump and lasing wavelengths shown in Extended Data Fig.~2. For the pump mode, we extracted an intrinsic Q-factor of 122,000, corresponding to a large absorption coefficient of $5.1\, \mathrm{dB\,cm^{-1}}$. At the lasing wavelength, the device exhibits an intrinsic Q-factor of 2.5 million and a loaded Q-factor of 1.5 million, corresponding to an intrinsic loss of $0.21\, \mathrm{dB\,cm^{-1}}$ and a loaded loss of $0.34\,\mathrm{dB\,cm^{-1}}$. Since lasing occurs when the optical gain exceeds the round-trip cavity loss, we estimate, based on the measured loaded Q, a lasing threshold power as low as \SI{1.6}{\micro\watt} (Supplementary Information). The experimental setup is illustrated in Extended Data Fig.~2a. The lasing output is then separated to the \SI{1.064}{\micro\meter} port of a fiber-based WDM and characterized with a power detector. Figure~\ref{fig3}a shows the continuous-wave light–light curve of a representative device featuring a ring radius of \SI{275}{\micro\meter} and width of \SI{2}{\micro\meter}, combined with a wrap waveguide ($w=\SI{1.31}{\micro\meter}$) and an extraction waveguide ($w=\SI{1.47}{\micro\meter}$), both having gap sizes of 150 nm. This device exhibits a lasing threshold of \SI{6}{\micro\watt}, while the minimum measured threshold of \SI{2.9}{\micro\watt} is achieved in a device with a \SI{200}{\nano\meter} extraction gap as shown in Supplementary Fig.~\changemarker{4}.

In the high-efficiency configuration, we switch to a single-port pulley-coupler design, where the extended interaction length effectively enhances signal extraction. A wider wrap waveguide is employed to support both pump and signal modes without signal cutoff. The wrap width is tuned between the individual phase-matching conditions of the pump and signal, with an optimal value of \SI{1.38}{\micro\meter} identified to provide strong coupling at both 808 nm and 1064 nm. Benefiting from efficient pump injection and signal extraction within a single coupler, the device achieves a 34\% slope efficiency from a single output port (Fig.~\ref{fig3}b).

To further examine the longitudinal lasing characteristics, the output was analyzed using either a calibrated optical spectrum analyzer (OSA) or a heterodyne spectrometer, as described in the Supplementary Information. The heterodyne spectrometer offers significantly higher spectral resolution, enabled by a swept-frequency tunable laser with sub-megahertz linewidths~\cite{HUI2009129}. In this configuration, a tunable laser (Newport TLB-6721) serves as the local oscillator and is heterodyned with the lasing output. The local oscillator wavelength is linearly swept from \SI{1.060}{\micro\meter} to \SI{1.070}{\micro\meter}, encompassing the entire emission band of the Nd:YAG $^4F_{3/2}$ transition. Figure~\ref{fig3}c presents the evolution of the lasing spectra with increasing on-chip pump power. At $P_{\text{pump}}=\SI{24}{\milli\watt}$, OSA measurement (left panel) reveals a broadened emission profile with a full width at half maximum of approximately 0.26 nm, limited by the OSA’s resolution bandwidth. The spectral baseline exhibits no resolvable substructure, obscuring any possible Fabry–Pérot parasitic modes induced by facet reflections~\cite{singh2025sub2w, liu2022photonic, brown1978parasitic}. In contrast, the heterodyne spectrum (right panel) reveals a distinct, spectrally sharp peak that remains stable across the entire sweep range, unambiguously confirming single-longitudinal-mode lasing. The intrinsic linewidth of the laser, further determined from the \changemarker{delayed self-heterodyne measurements, is 16.0 kHz (Supplementary Information).}

\vspace{0.6cm}
\noindent\textbf{$\upmu$-MOPA: Performance} 

\begin{figure*}[t]
\centering
\includegraphics[trim={0cm 0cm 0cm 0cm},clip,width=1\linewidth]{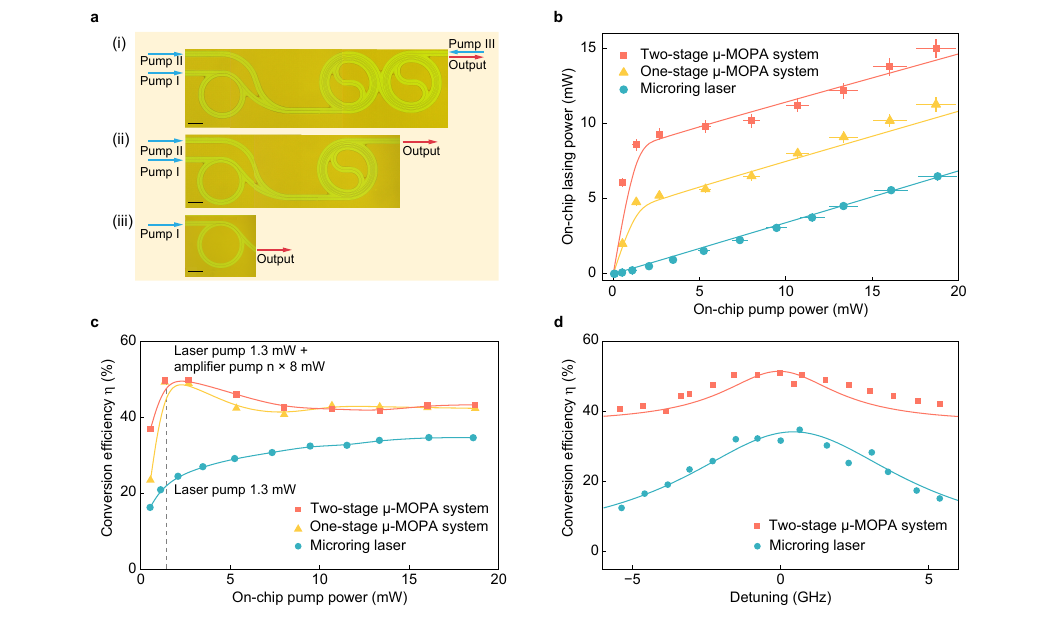}
\caption{
{\fontseries{bx}\selectfont $\upmu$-MOPA Nd:YAG laser. }
{\fontseries{bx}\selectfont a,} Optical microscope image of the integrated devices, showing the cascaded amplifier stages pumped by separate input waveguides. Scale bar: \SI{200}{\micro\meter}.
{\fontseries{bx}\selectfont b,} L–L characteristics of two-stage $\upmu$-MOPA, single-stage $\upmu$-MOPA, and standalone microring laser. Solid lines: numerical model; points: measurements.
{\fontseries{bx}\selectfont c,} Conversion efficiency ($\eta$) versus microring laser pump power for the two-stage $\upmu$-MOPA and microring laser. The total pump power corresponds to Pump I for the microring laser and Pump I + II + III for the two-stage $\upmu$-MOPA. At a Pump I power of 1.3 mW, the two-stage $\upmu$-MOPA achieves a conversion efficiency exceeding 50\% due to effective amplification.
{\fontseries{bx}\selectfont d,} Conversion efficiency as a function of detuning at Pump I = 1.3 mW, showing enhanced spectral stability of the two-stage $\upmu$-MOPA compared with the microring laser.
}
\label{fig4}
\end{figure*}

Benefiting from the high-gain waveguide amplifier and the low-threshold single-mode laser demonstrated above, we next present a high-efficiency integrated Nd:YAG laser based on the $\upmu$-MOPA architecture. By decoupling signal generation from power amplification, this configuration enables high output power scaling while maintaining single-mode optical coherence. Figure~\ref{fig4}a shows optical images of (i) a two-stage $\upmu$-MOPA system, (ii) a single-stage $\upmu$-MOPA system, and (iii) an individual microring laser. A pulley-coupled microring resonator serves as the master oscillator, using the same coupling parameters as in Fig.~\ref{fig3}b. The extracted seed is routed through a directional coupler that separates the pump and signal paths, directing the \SI{1.064}{\micro\meter} signal into the top arm while leaving residual pump in the lower waveguide. Additional \SI{0.808}{\micro\meter} pump inputs are introduced through the top waveguides, where they combine with the seed for subsequent amplification. A lensed fiber array is used for pump simultaneously coupling, which exhibits an average insertion loss of approximately 6.9 dB per channel. In the two-stage $\upmu$-MOPA system, Pump III is introduced in the counter-propagating direction to simplify the design. This configuration maintains performance equivalent to the co-propagating case (Supplementary Fig.~\changemarker{6}) once the amplified signal reaches the high power regime, where gain dynamics are governed primarily by the strong seed signal rather than by ASE~\cite{Singh2025WattAmplifier}. Notably, the amplifier pumps need not be single-mode nor precisely resonant with microring laser pump source.

Figure~\ref{fig4}b presents the light-light characteristics for the three laser layouts. The on-chip microring laser pump power (Pump I) is varied along the x-axis, while the amplifier pump powers (Pump II and III) are fixed at 8 mW on-chip power per channel. The standalone microring laser exhibits constrained output power, primarily limited by suboptimal coupling conditions that preclude simultaneous optimization of pump and signal coupling, as well as by bidirectional lasing, which restricts power extraction to a single propagation direction. In contrast, both single-stage and two-stage $\upmu$-MOPA systems demonstrate substantially enhanced output performance owing to efficient pump-to-signal conversion in the amplifier stage. The output rises steeply in the small-signal gain regime and gradually levels off as the amplifier gain saturates. Numerical simulations (solid lines) closely match the experimental data, validating the model accuracy. The two-stage $\upmu$-MOPA achieves a maximum lasing output of 15 mW, representing a threefold enhancement over the standalone laser. Moreover, leveraging the modular $\upmu$-MOPA configuration, the output power can be further boosted by cascading additional amplifier stages. Simulations (Supplementary Fig.~\changemarker{7}) predict an output exceeding 40 mW with eight amplifier stages at 8 mW pump power per amplifier, and up to 100 mW when the amplifier pump power is increased to 20 mW for each stage. Such enhancements could be achieved in the future by incorporating taper couplers, which enable low-loss butt-coupling with commercial laser diodes~\cite{Jia2023edgecoupler,He2019bilayercoupler,siddharth2025ultrafast}.

Figure ~\ref{fig4}c summarizes the conversion efficiency ($\eta$) as a function of total on-chip pump power, defined as $P_{\text{total}}$ = Pump I for the microring laser and $P_{\text{total}}$ = Pump I + II + III for the two-stage $\upmu$-MOPA system. \changemarker{The measured on-chip efficiency isolates the intrinsic performance of the gain architecture, enabling direct comparison to the quantum-defect limit independent of external coupling losses and diode wall-plug efficiency.} The microring laser exhibits a maximum $\eta$ of approximately 34\%, beyond which the efficiency saturates due to incomplete pump utilization and signal extraction. In contrast, the two-stage $\upmu$-MOPA shows a peak efficiency exceeding 50\% when the seed laser is pumped at 1.3 mW and each amplifier receives 8 mW. \changemarker{Under these operating conditions, the overall efficiency is largely determined by the amplifier stages, where most of the pump power is delivered. In practice, the maximum achievable system efficiency is constrained by waveguide loss at the signal wavelength, incomplete pump absorption, and ~1.3 dB pump insertion loss from the directional coupler.} By further optimizing pump power distribution between the microring laser and the amplifiers, the overall conversion efficiency of the $\upmu$-MOPA can reach up to \changemarker{68}\%, with the amplifier itself exhibiting an $\eta$ of approximately 72\%, approaching the theoretical quantum limit (Supplementary Information). 
As shown in Fig.~\ref{fig4}d, the $\upmu$-MOPA system maintains high conversion efficiency over a broad range of pump-wavelength detuning. Within a $\pm~$5 GHz range, its efficiency decreases by only 23\%, in contrast to a 64\% drop observed for the microring laser alone. This highlights the robustness of the $\upmu$-MOPA architecture, which decouples gain and resonance conditions, thereby stabilizing lasing performance against detuning and fabrication variations. \changemarker{The noise characteristics of the µ-MOPA system are presented in the Supplementary  Fig.~10, with delayed self-heterodyne measurements showing no observable degradation after amplification and preservation of the intrinsic linewidth.}

In conclusion, we have demonstrated a photonic-integrated Nd:YAG master-oscillator–power-amplifier system that decouples microwatt-level seed generation from efficient on-chip power scaling. A double-resonant microring produces a single-mode seed with thresholds down to \SI{2.9}{\micro\watt}, while a co-integrated Nd:YAG waveguide amplifier provides up to \SI{46.6}{\decibel} net gain in a \SI{2.3}{\centi\meter} spiral using only \SI{12}{\milli\watt} of pump. Combining these elements, the integrated $\upmu$-MOPA achieves high continuous-wave output power (up to \SI{15}{\milli\watt} in a two-stage configuration) while preserving spectral purity.  Separating the oscillator and amplifier removes the inherent trade-off between pump utilization and signal extraction in single-ring designs, while ensuring robust operation across coupling conditions and pump detuning. \changemarker{The bonded Nd:YAG layer shows no measurable degradation in threshold, slope efficiency, or gain under repeated operation.}

\changemarker{While the efficiencies reported here are defined with respect to the optical pump power coupled into the chip, the overall system-level efficiency is currently limited by fiber-to-chip coupling losses and the wall-plug efficiency of external pump diodes. Reducing coupling losses through improved mode converters and packaging, together with co-packaged or integrated pump sources, will be critical for approaching high electrical-to-optical efficiency in fully integrated systems.} \changemarker{In parallel with crystalline gain integration, rare-earth--doped amorphous thin-film platforms such as Al$_2$O$_3$-on-silicon~\cite{mu2020high, yang2010high,jongebloed2026fiber} have demonstrated scalable on-chip gain. Compared to these systems, crystalline gain media such as Nd:YAG offer higher emission cross-sections and thermal conductivity; here, heterogeneous integration combines the best of both worlds, low-loss waveguides and high-performance crystalline gain, for improved efficiency and power handling.} 

\changemarker{Extension of this approach to other gain materials depends on both optical and mechanical properties. Mechanically robust crystals such as YAG and sapphire are well suited for direct bonding and high-power operation, whereas softer materials (e.g., fluoride crystals~\cite{metz2014high, atherton1993oxide}) may require modified bonding strategies, such as intermediate layers or reduced thermal budgets. These considerations suggest that the $\upmu$-MOPA architecture can be extended to a broad class of materials with appropriate integration engineering.}

The $\upmu$-MOPA platform is inherently scalable: higher-power operation is accessible through optimized pump distribution, additional amplifier stages, and \changemarker{reduced facet and coupling losses}. Hybrid integration of higher-brightness pump diodes, combined with enhanced thermal management, \changemarker{can} further improve pump delivery efficiency and support additional power scaling. Beyond continuous-wave output, leveraging the Kerr nonlinearity of the Si$_3$N$_4$ platform~\cite{Qiu2025Mamyshev} or integrating fast modulators~\cite{Guo2023MLL} could extend this Nd:YAG–Si$_3$N$_4$ engine to tunable, narrow-linewidth or pulsed operation, with immediate relevance to chip-scale clocks, quantum networks, coherent LiDAR, and precision optical systems.

\vspace{1cm}
\noindent\textbf{Methods} \\[4mm]
\noindent\textbf{1. Device fabrication.}\\
The SiN photonic circuit patterns and bonding regions are defined by electron-beam lithography using FOx-16 (Dow-Corning) as the etching mask. The Si$_3$N$_4$ layer is then etched with a fluorine-based inductively coupled plasma reactive ion etching process~\cite{wang2024heterogeneous}, followed by mask removal using buffered oxide etch. To ensure high surface quality prior to bonding, the commercial Nd:YAG crystal is polished with a silica slurry through chemical–mechanical polishing. Both the Nd:YAG and SiN chips are sequentially cleaned with acetone and isopropyl alcohol, and their surfaces are activated using piranha solution and oxygen plasma. The two chips are subsequently flip-chip bonded at room temperature. The contact area between Nd:YAG and Si$_3$N$_4$ is estimated to be approximately 80\%, with an RMS surface roughness of 0.3 nm on both surfaces, resulting in strong and reliable bonding. The completed chip is finally cleaved along the silicon 100-plane to enable side coupling.

\vspace{3mm}
\noindent\textbf{2. Characterization of the waveguide amplifier.}\\
On-chip Nd:YAG amplifier characterization is performed using the setup shown in Extended Data Fig.~1a. The 808-nm pump is provided by a commercial Ti:sapphire laser (SolsTiS, M Squared), while a laser diode (Aerodiode, 1064LD-2-0-0) centered at \SI{1.064}{\micro\meter} serves as the signal source. The pump and signal are combined through separate input ports of a fiber-based WDM and coupled into the device. The amplified output is collected via a lensed fiber from the extraction waveguide and directed to another fiber-based WDM, which separates the residual pump from the signal output. For small-signal gain characterization, an optical spectrum analyzer (Yokogawa AQ6370D) is used to measure the output spectra with and without the pump. The passive waveguide loss is calibrated in this process through statistical measurements of waveguides with different lengths, yielding an average loss of approximately 1 dB/cm for the 2.3-cm-long amplifier.

\vspace{3mm}
\noindent\textbf{3. Integrated microring lasers measurement.}\\
Laser characterization is conducted using the setup shown in Extended Data Fig.~2a, employing the same pump light sources as used for the waveguide amplifier measurements. A fiber-based variable optical attenuator (VOA) is placed before coupling light into the chip, enabling linear control of the pump power for recording the light–light curve. For the heterodyne spectrum analyzer measurement, a commercial tunable diode laser (Newport TLB-6721) is used as the local oscillator and heterodyned with the lasing signal through a 3-dB coupler. The two outputs of the coupler are detected by a balanced photodiode pair (PDB471C), effectively suppressing direct detection intensity noise in the heterodyne coherent receiver. The resulting RF signal is then passed through a 1 MHz bandwidth electrical bandpass filter and recorded using a power monitor (Mini-Circuits ZX47-60-S+). The heterodyne linewidth is measured using the same local oscillator. The beat signal is detected by a high-speed photodetector (New Focus 1611) and analyzed with an electrical spectrum analyzer (Agilent N9020A, resolution bandwidth = 120 kHz).

\vspace{3mm}
\noindent\textbf{4. $\upmu$-MOPA laser measurement.}\\
Single-mode laser diodes (Aerodiode 808LD-1-0-0, butterfly package) are used as the three pump sources. Each diode is driven and temperature-stabilized using a compact LD and temperature controller (Thorlabs CLD1015). The emission wavelength of the diodes can be tuned via temperature adjustment, allowing precise alignment of the pump light to the microring cavity resonance. The diode outputs are coupled into the chip through separate channels of a commercial lensed fiber array, which is mounted on a five-axis stage to ensure precise alignment across all channels. For the $\upmu$-MOPA stability measurements, a tunable commercial Ti:sapphire laser (M2 SolsTiS) is employed as the microring pump source (Pump I) to investigate the dependence of the $\upmu$-MOPA output on wavelength detuning.

\vspace{4 mm}
\noindent\textbf{Data availability} 
The data that support the findings of this study are available from the corresponding authors upon reasonable request.

\vspace{4 mm}
\noindent\textbf{Code availability} 
All relevant computer codes supporting this study are available from the corresponding author upon reasonable request.

\vspace{4 mm}
\noindent \textbf{Acknowledgements.} This work was supported by DARPA under contract No. HR0011-20-2-0045. We thank our cleanroom staff Yong Sun, Lauren McCabe, Yeongjae Shin, Kelly Woods and Michael Rooks for assistance with device fabrication. 

\vspace{4 mm}
\noindent \textbf{Author contributions:} Y.G., Y.W., and H.T. conceived the experiment. Y.G. and Y.W. designed the devices. Y.G. fabricated the devices and performed the measurements. Y.G. and Y.W. analyzed the data. H.Z. and Y.W. performed the simulation. All authors contributed to writing the manuscript. H.T. supervised the work.

\vspace{4 mm}
\noindent \textbf{Competing interests.} The authors declare no competing interests.
\bibliographystyle{naturemag}
\newpage
\def\bibsection{\section{\textbf{references}}}

\newpage

\begin{center}
\centering
\includegraphics[width=\linewidth,trim={0cm 0cm 0cm 0cm},clip]{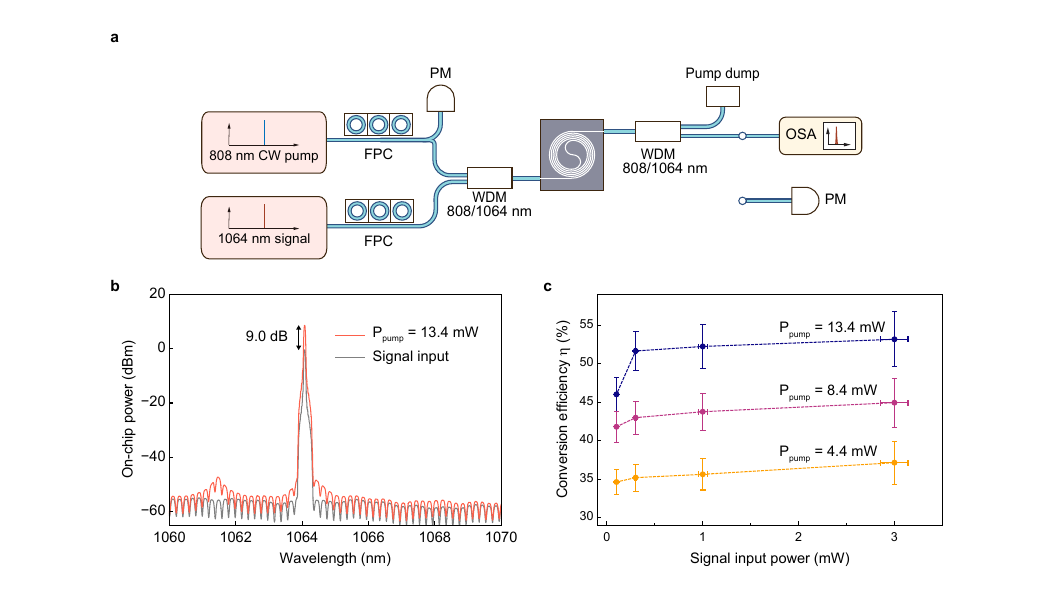}
\begin{flushleft}
\textbf{Extended Data Fig.~1.}
\textbf{a,} Experimental setup for amplifier characterization.
\textbf{b,} Spectra of large-signal amplification with a net gain of \changemarker{9.0} dB. \changemarker{The periodic oscillations arise from WDM-induced etalon effects rather than on-chip facet reflections, as verified by an off-chip control measurement.
\textbf{c,} Conversion efficiency $\eta$ as a function of on-chip input signal power for different pump levels. A maximum efficiency of 53.2\% is achieved at 13.4 mW pump power and 3 mW input signal power. Error bars indicate coupling-loss uncertainty.}
\end{flushleft}
\label{ExtendedFig1}
\end{center}

\newpage
\begin{center}
\centering
\includegraphics[trim={0cm 0cm 0cm 0cm},clip,width=1\linewidth]{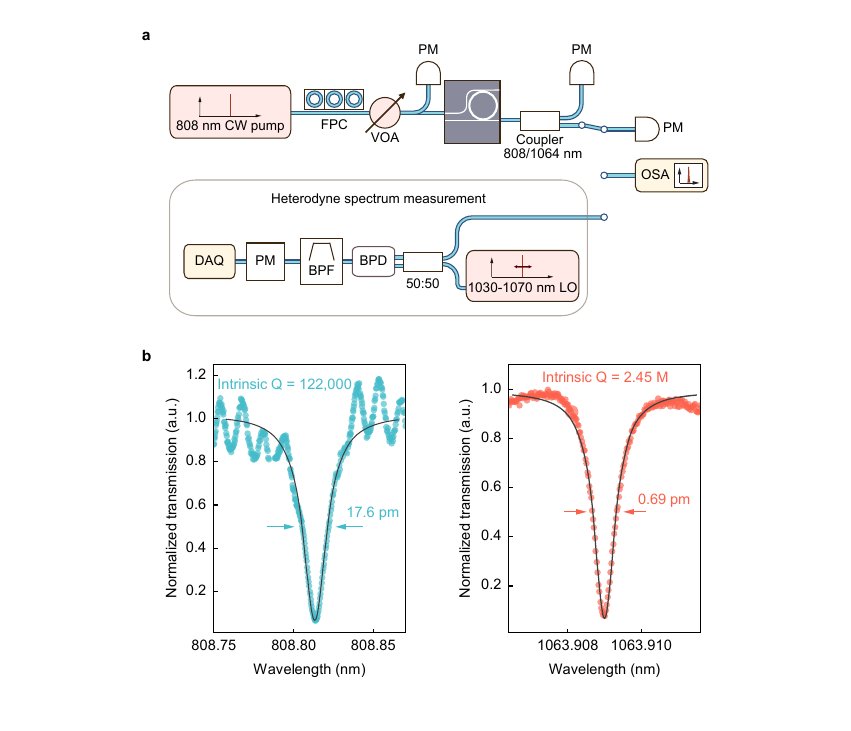}
\begin{flushleft}
\textbf{Extended Data Fig.~2.}
{\fontseries{bx}\selectfont Characterization of the standalone Nd:YAG microring laser.}
{\fontseries{bx}\selectfont a,} Experimental setup for power and spectrum measurement. A tunable laser (Newport TLB-6721) serves as the local oscillator and is heterodyned with the lasing output using a 3-dB coupler and balanced photodiode pair to suppress intensity noise. The resulting RF signal is filtered with a 1-MHz-bandwidth band-pass electrical filter.
{\fontseries{bx}\selectfont b,} Measured optical Q-factors at the pump and lasing wavelengths.
\end{flushleft}
\label{ExtendedFig2}
\end{center}

\clearpage


\clearpage
\appendix
\onecolumngrid
\nolinenumbers



\clearpage
\onecolumngrid

\setstretch{0.92}

\setcounter{page}{1}

\providecommand{\theHpage}{\arabic{page}}
\renewcommand{\theHpage}{S\arabic{page}}

\setcounter{section}{0}
\setcounter{subsection}{0}
\setcounter{figure}{0}
\setcounter{table}{0}
\setcounter{equation}{0}

\renewcommand{\thesection}{\Roman{section}}
\renewcommand{\thesubsection}{\Alph{subsection}}

\renewcommand{\thefigure}{S\arabic{figure}}
\renewcommand{\thetable}{S\arabic{table}}
\renewcommand{\theequation}{S\arabic{equation}}

\renewcommand{\theHfigure}{S\arabic{figure}}
\renewcommand{\theHtable}{S\arabic{table}}
\renewcommand{\theHequation}{S\arabic{equation}}
\renewcommand{\theHsection}{S\Roman{section}}
\renewcommand{\theHsubsection}{S\Roman{section}.\Alph{subsection}}

\counterwithout*{equation}{section}
\counterwithout*{equation}{subsection}

\renewcommand{\theequation}{S\arabic{equation}}

\begin{center}
{\LARGE\bfseries Supplementary Information}
\end{center}

\vspace{0.5cm}

\section*{Contents}

\begingroup
\newcommand{\SIContentLine}[2]{%
  \noindent\makebox[\textwidth][l]{\hyperref[#1]{#2}\hfill\hyperref[#1]{\pageref*{#1}}}\par
}
\newcommand{\SISpace}{\vspace{0.9em}}
\SIContentLine{si:note1}{I.\quad Photonic chip design and characteristics}
\SIContentLine{si:note1a}{\hspace*{1.6em}A.\quad Optical confinement and modal overlap between pump and signal}
\SIContentLine{si:note1b}{\hspace*{1.6em}B.\quad Directional coupler wavelength-division multiplexer}
\SIContentLine{si:note1c}{\hspace*{1.6em}C.\quad Heterodyne spectrum measurement of amplifier output}
\SIContentLine{si:note1d}{\hspace*{1.6em}D.\quad Microring laser threshold}
\SIContentLine{si:note1e}{\hspace*{1.6em}E.\quad Lasing spectrum of $\upmu$-MOPA devices}
\SISpace
\SIContentLine{si:note2}{II.\quad Modeling of the Integrated Nd:YAG Amplifier and Laser}
\SISpace
\SIContentLine{si:note3}{III.\quad Heterodyne spectrum measurement}
\SISpace
\SIContentLine{si:note4}{IV.\quad Noise characterization of single-ring and $\upmu$-MOPA lasers}
\SISpace
\SIContentLine{si:references}{V.\quad References}
\endgroup

\clearpage


\section{Photonic chip design and characteristics}\label{si:note1}
\subsection{Optical confinement and modal overlap between pump and signal}\label{si:note1a}

The simulated transverse electric (TE$_{00}$) mode profiles of the pump and signal in the microring laser and amplifier are shown in Fig.~S1a, obtained using the Lumerical MODE solver. The SiN waveguide, with a height of 380 nm and a width of 2 µm, was designed to minimize sidewall-induced scattering losses while maintaining strong optical confinement. The calculated effective mode areas are approximately 0.82 µm² for the pump and 1.1 µm² for the signal, resulting in a pump–signal modal overlap exceeding 92\%. 

\begin{center}
\centering
\includegraphics[width=\linewidth,trim={0cm 0cm 0cm 0cm},clip]{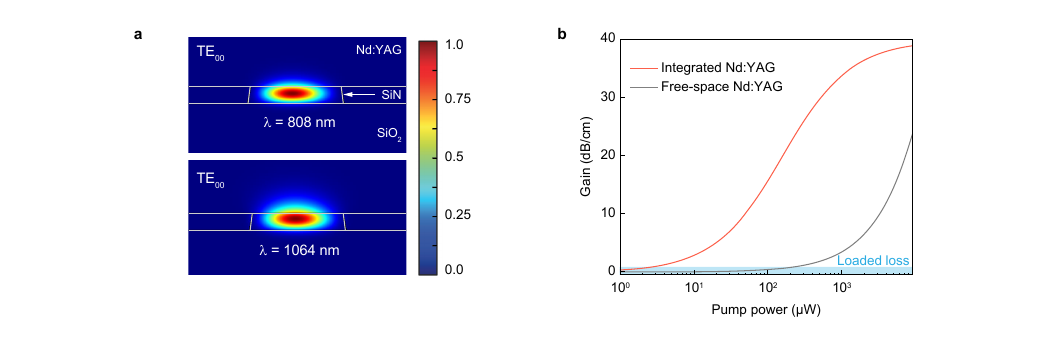}
\begin{flushleft}
\textbf{Fig. S1. Pump/signal modes and gain calculation. }
\textbf{a}, Simulated mode intensity profile at the pump (808 nm; top) and lasing (1064 nm; bottom) wavelength, with modal intensity overlap $\Gamma_{p,s}$ of 92\%.
\textbf{b}, Simulated optical gain as a function of pump power for integrated and free-space Nd:YAG lasers.
\end{flushleft}
\label{Fig.S1}
\end{center}

To quantitatively evaluate the optical gain, we calculated the gain as a function of intracavity pump power for both the integrated and free-space Nd:YAG lasers. The free-space configuration was assumed to have an effective mode area of 200~µm\textsuperscript{2}~\cite{wang2023photonic}. Under steady-state pumping conditions ($I_s = W_s = 0$) and using the population inversion model from Ref.~\cite{Liu2005PhotonicDevices},  the fraction of excited Nd\textsuperscript{3+} ions is given by

\begin{equation}
\frac{N_3}{N_{\mathrm{tot}}} = \frac{W_p}{\frac{1}{\tau_3} + W_p}
= \frac{\dfrac{P_p \sigma_p}{A_{\mathrm{eff}} h \nu_p}}{\frac{1}{\tau_3} + \dfrac{P_p \sigma_p}{A_{\mathrm{eff}} h \nu_p}},
\end{equation}
where $P_p$ is the pump power, $A_{\mathrm{eff}}$ is the effective mode area, $\sigma_p$ is the pump absorption cross-section, $h \nu_p$ is the pump photon energy, and $\tau_3$ is the upper-level lifetime.  

The optical gain is expressed as
\changemarker{
\begin{equation}
g_0 = N_3 \, \sigma_e \, \Gamma_p \, \delta_{sp}\,
\end{equation}}
where \changemarker{$\sigma_e$ is the emission cross-section,} $\delta_{sp}$ represents the pump-signal modal overlap factor and $\Gamma_p$ denotes the overlap factor between the pump mode and the gain medium. For the 380-nm-thick SiN waveguide used in the integrated device, $\Gamma_p = 15\%$.

As shown in Fig.~S1b, the photonic-circuit-integrated Nd:YAG laser exhibits an optical gain nearly one order of magnitude higher than that of its free-space counterpart and a lasing threshold of \SI{1.6}{\micro\watt} based on a loaded loss of $0.34\,\mathrm{dB\,cm^{-1}}$ of the SiN platform. This enhancement originates from the strong optical confinement within the waveguide, which enables efficient pump absorption and rapid population inversion even under sub-milliwatt pump levels. A full modeling description, including the ASE formulation, is provided in Section II.

\subsection{Directional coupler wavelength-division multiplexer}\label{si:note1b}

To introduce an additional pump into the amplifier stage, we implement a wavelength-selective directional coupler that functions as an on-chip wavelength-division multiplexer (WDM), enabling simultaneous routing of the 808-nm pump and 1064-nm signal~\cite{Wang1992coupler,Thottoli2023coupler}. As the two wavelengths exhibit different confinement factors and coupling strengths in the same waveguide, the signal experiences higher bending loss and a shorter coupling length than the pump. The coupler design shown in Fig. S2a consists of two parallel waveguides with a coupling length \textit{L} = 360 µm and a gap \textit{g} = 0.3 µm.  It is designed so that only the longer-wavelength signal meets the phase-matching condition and transfers to the adjacent waveguide, while the pump remains in its original path. This configuration ensures that the signal propagates primarily along straight waveguide sections, minimizing bending-induced loss.

Simulations of the power distributions (Fig. S2b) verify the designed wavelength-dependent routing, yielding output transmission efficiencies of 85\% at 808 nm and 97\% at 1064 nm. Increasing the coupling gap would raise the pump transmission, but at the cost of a much longer coupling length. This option is not used to maintain a compact footprint.

\begin{center}
\centering
\includegraphics[width=\linewidth,trim={0cm 0cm 0cm 0cm},clip]{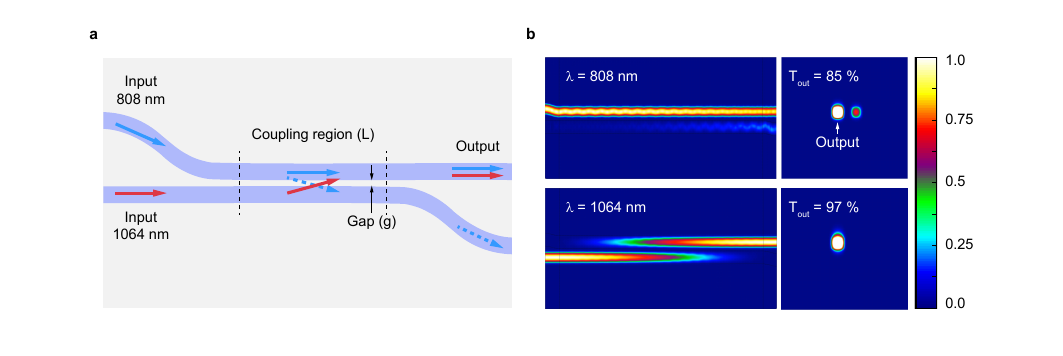}
\begin{flushleft}
\textbf{Fig. S2. Directional coupler for pump–signal multiplexing. }
\textbf{a}, Schematic of the directional coupler used to combine the 808-nm pump and the 1064-nm lasing signal. The structure is configured to ensure that the pump mode is retained while routing the signal mode into the upper waveguide.
\textbf{b}, Power distributions of the wavelength-division multiplexer for TE modes for both wavelengths. At the output port, the transmission efficiencies are 
$T_{\mathrm{out}} = 85\%$ at 808 nm and $97\%$ at 1064 nm.
\end{flushleft}
\label{Fig.S1}
\end{center}

\newpage

\changemarker{\subsection{Heterodyne spectrum measurement of amplifier output}}\label{si:note1c}

\changemarker{As the secondary transition between different Stark levels of \( ^4F_{3/2} \rightarrow {}^4I_{11/2} \) can support laser oscillation near \SI{1061.5}{\nano\meter}, as well documented in the literature (e.g., Ref.~\cite{Cho2013NdYAG}), we performed a heterodyne spectrum measurement to rule out the possibility of parasitic lasing at \SI{1061.5}{\nano\meter}, associated with the spectral feature observed in Fig.~2b. The measurement setup is shown in Fig.~S3a. A tunable laser (Newport TLB-6721) was used as the local oscillator and heterodyned with the amplified output signal. As shown in Fig.~S3b, a single narrow beatnote is observed at \SI{1064}{\nano\meter}, with no additional beatnotes detected across the entire sweep range. This confirms that the spectral feature at \SI{1061.5}{\nano\meter} arises from amplified spontaneous emission (ASE), rather than parasitic lasing.}

\changemarker{
\begin{center}
\centering
\includegraphics[width=\linewidth,trim={0cm 0cm 0cm 0cm},clip]{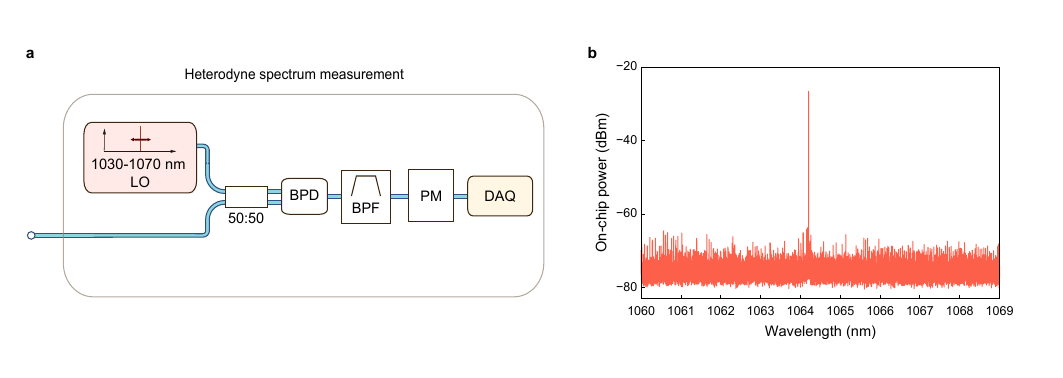}
\begin{flushleft}
\textbf{Fig. S3. Heterodyne beatnote of amplifier output. }
\textbf{a}, Schematic of the experimental setup used for heterodyne measurement of the amplified lasing signal.
\textbf{b}, Measured heterodyne spectrum showing a single narrow beatnote at 1064 nm, with no additional features near 1061.5 nm, confirming the absence of parasitic lasing.
\end{flushleft}
\label{Fig.S3}
\end{center}
}

\subsection{Microring laser threshold}\label{si:note1d}

\begin{center}
\centering
\includegraphics[width=\linewidth,trim={0cm 0cm 0cm 0cm},clip]{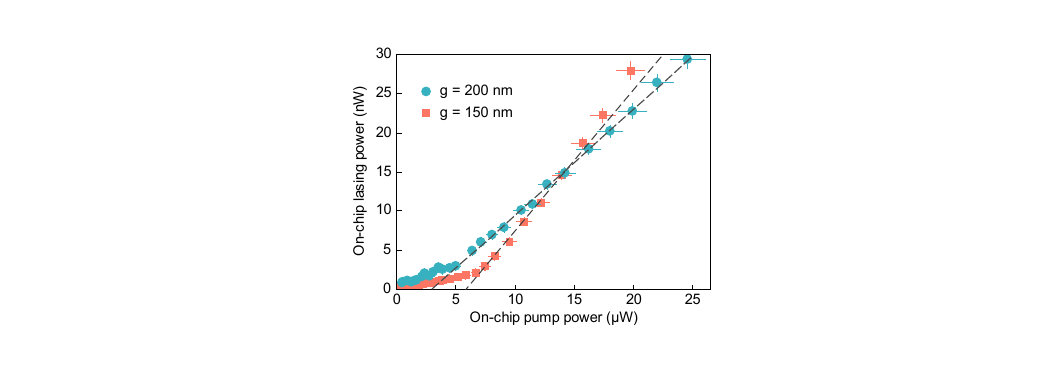}
\begin{flushleft}
\textbf{Fig. S4. Microring lasing threshold. }
Light–light curve near threshold for microring lasers with extraction gaps of 150 nm (orange) and 200 nm (blue).
\end{flushleft}
\label{Fig.S4}
\end{center}

Supplementary Fig.~4 shows the light–light curves for microring lasers employing a two-port coupling scheme with varying extraction-gap sizes. The device with a \SI{200}{\nano\meter} gap exhibits the lowest measured threshold of \SI{2.9}{\micro\watt}.

\changemarker{\subsection{Lasing spectrum of $\upmu$-MOPA devices}}\label{si:note1e}

\changemarker{The lasing spectra of the microring laser and the one- and two-stage $\upmu$-MOPA devices are shown in Fig.~S5a. The $\upmu$-MOPA devices maintain single-mode lasing centered at 1064.2 nm, consistent with the microring laser after the addition of amplification stages. Supplementary Fig.~5b shows the lasing output power (in mW) over a $\pm~$ 0.2 nm window around the lasing wavelength, exhibiting a progressive increase from the microring laser to the one- and two-stage $\upmu$-MOPA configurations. A slight red shift ($\sim$ 0.01 nm) of the lasing peak is observed with increasing pump power, attributed to thermal effects. These results demonstrate that the $\upmu$-MOPA architecture preserves spectral purity while enabling substantial power scaling.}
\changemarker{
\begin{center}
\centering
\includegraphics[width=\linewidth,trim={0cm 0cm 0cm 0cm},clip]{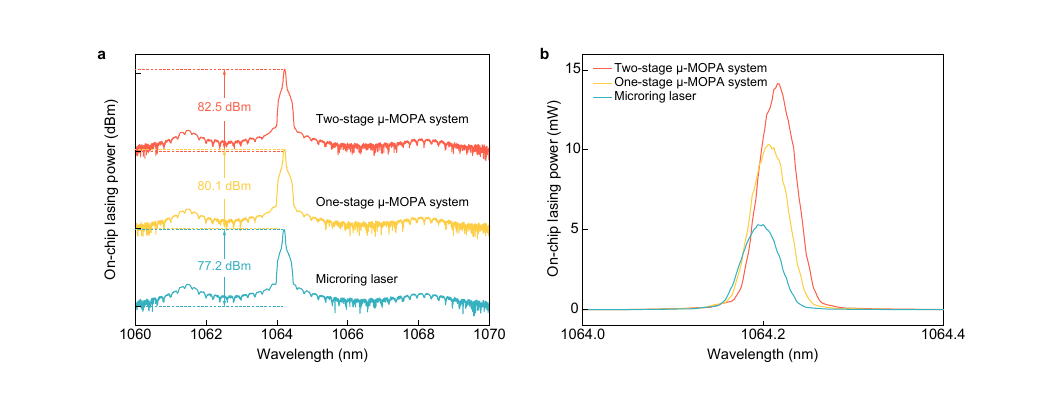}
\begin{flushleft}
\textbf{Fig.~S5. Lasing spectrum of $\upmu$-MOPA systems. }
\textbf{a}, Lasing spectra of the microring laser, one-stage $\upmu$-MOPA, and two-stage $\upmu$-MOPA systems, with traces vertically offset by 80 dB per division for clarity.
\textbf{b}, Corresponding lasing output power (mW) over a $\pm$ 0.2 nm window around the lasing wavelength. The on-chip power is shown in a linear scale to highlight the progressive power scaling from the microring laser to the one- and two-stage $\upmu$-MOPA configurations. 
\end{flushleft}
\label{Fig.S4}
\end{center}
}

\section{Modeling of the Integrated Nd:YAG Amplifier and Laser}\label{si:note2}
We develop a quantitative model to describe the behavior of the integrated Nd:YAG amplifier and laser. Nd:YAG operates as a four-level laser system, as illustrated in Fig.~1a. For a given spatial position \((x, y)\) within the cross-section of the Nd:YAG crystal, the population dynamics of the four energy levels follow the rate equations:

\begin{align}
\frac{dN_4}{dt} &= W_{14} N_1 - W_{41} N_4 - \frac{N_4}{\tau_4}, \label{eq3} \\[6pt] 
\frac{dN_3}{dt} &= \frac{N_4}{\tau_4} - \frac{N_3}{\tau_3} 
- \frac{I_s}{h\nu_s}\left(N_3\sigma_e^s - N_2\sigma_a^s\right), \\[6pt] 
\frac{dN_2}{dt} &= -\frac{N_2}{\tau_2} 
+ \frac{I_s}{h\nu_s}\left(N_3\sigma_e^s - N_2\sigma_a^s\right), \\[6pt] 
\frac{dN_1}{dt} &= W_{41} N_4 - W_{14} N_1 + \frac{N_2}{\tau_2}. \label{eq6}
\end{align}
The pump-induced transition rates are given by
\[
W_{14} = \frac{I_p}{h\nu_p}\sigma_a^p, \quad
W_{41}=\frac{g_1}{g_4}W_{14}. 
\]

Here, \( N_i \) and \( \tau_i \) denote the population density and lifetime of energy level \(|i\rangle\), respectively. \( I_s \) and \( I_p \) are the signal and pump intensities at position \((x, y)\). The parameters \( \sigma_e \) and \( \sigma_a \) represent the emission and absorption cross-sections for the pump (\(p\)) and signal (\(s\)) transitions. The relevant parameters are:
\[
\sigma_a^p = 7.7 \times 10^{-20}\,\mathrm{cm^2}, \quad
\sigma_e^p = 3.85 \times 10^{-20}\,\mathrm{cm^2}, \quad
\sigma_a^s = \sigma_e^s = 2.8 \times 10^{-19}\,\mathrm{cm^2}.
\]
The degeneracies of the ground and upper pump levels are \( g_1 = 1 \) and \( g_4 = 2 \), respectively. The total ion concentration is
\[
N_{\text{tot}} = N_1 + N_2 + N_3 + N_4 = 1.518 \times 10^{26}\,\mathrm{m^{-3}}.
\]

Under steady-state conditions, the population rates satisfy
\[
\frac{dN_i}{dt} = 0 \quad (i = 1, 2, 3, 4),
\]
and Eqs.~\ref{eq3}-\ref{eq6} can be solved simultaneously to yield closed-form expressions for the level populations:

\begin{align}
N_3 &= 
\frac{N_{\text{tot}}}{
1 + 
\frac{\Gamma_s\sigma_e^s}{1+\Gamma_s\sigma_a^s} +
\tau_4\!\left(
\frac{1}{\tau_3} + 
\frac{I_s\sigma_e^s}{h\nu_s(1+\Gamma_s\sigma_a^s)}
\right)
\!\left(
1 + \frac{W_{41}+1/\tau_4}{W_{14}}
\right)
}, \\[6pt]
N_2 &= 
\frac{\Gamma_s\sigma_e^s}{1+\Gamma_s\sigma_a^s}\, N_3, \\[6pt]
N_4 &=
\tau_4 N_3
\!\left(
\frac{1}{\tau_3} +
\frac{I_s\sigma_e^s}{h\nu_s(1+\Gamma_s\sigma_a^s)}
\right), \\[6pt]
N_1 &=
\frac{W_{41}+1/\tau_4}{W_{14}}\, N_4, 
\end{align}
where \( \Gamma_s = I_s \tau_2 / (h\nu_s) \).

From these populations, the local optical gain for the signal and the pump absorption at a given spatial position \((x, y)\) are given by:
\begin{equation}
g(x,y) = N_3(x,y)\sigma_e^s - N_2(x,y)\sigma_a^s, \qquad
\alpha(x,y) = N_1(x,y)\sigma_a^p - N_4(x,y)\sigma_e^p.  \label{eq11}
\end{equation}
To model the mode evolution within the integrated structure, the effective optical gain and absorption for the signal and pump modes must be evaluated based on their spatial field distributions. Using a first-order perturbation approach, the effective gain (or loss) coefficients are expressed as:
\[
g_{\text{eff}} =
\frac{\displaystyle\iint g(x,y) I_s(x,y)\, dS}
{\displaystyle\iint I_s(x,y)\, dS}, \qquad
\alpha_{\text{eff}} =
\frac{\displaystyle\iint \alpha(x,y) I_p(x,y)\, dS}
{\displaystyle\iint I_p(x,y)\, dS}. \tag{S12} \label{eq12}
\]
Here, \( I_s(x,y) \) and \( I_p(x,y) \) denote the normalized intensity distributions of the signal and pump modes, obtained from the simulations (Fig.~S1a). By combining Eqs.~\ref{eq11} and~\ref{eq12}, the effective optical gain for the signal and the pump absorption can be numerically determined for a given pair of pump and signal powers.

For simulating the integrated Nd:YAG amplifier, the spatial evolution of the pump and signal powers within the amplifier is governed by the following coupled equations~\cite{becker1999erbium}:
\begin{align}
\frac{dP_p}{dz} &= -\alpha_{\text{eff}} P_p - \alpha_p P_p, \tag{S13}\\[6pt]
\frac{dP_s}{dz} &= g_{\text{eff}} P_s - \alpha_s P_s, \tag{S14}\\[6pt]
\frac{dP_{\text{ASE}}^{+}}{dz} &= g_{\text{eff}} P_{\text{ASE}}^{+} - \alpha_s P_{\text{ASE}}^{+} + g_{\text{sp}} h\nu\Delta\nu, \tag{S15} \label{eq15}\\[6pt]
\frac{dP_{\text{ASE}}^{-}}{dz} &= -g_{\text{eff}} P_{\text{ASE}}^{-} + \alpha_s P_{\text{ASE}}^{-} - g_{\text{sp}} h\nu\Delta\nu. \tag{S16} \label{eq16}
\end{align}
where \[
g_{\text{sp}} =
\frac{\displaystyle\iint N_3(x,y)\sigma_e^s I_s(x,y)\, dS}
{\displaystyle\iint I_s(x,y)\, dS}, 
\]
Here, \( P_p \) and \( P_s \) denote the pump and signal powers at position \( z \) along the propagation axis. \( \alpha_p \) and \( \alpha_s \) are the intrinsic waveguide scattering losses for the pump and signal modes, respectively. \( P_{\text{ASE}}^{+} \) and \( P_{\text{ASE}}^{-} \) represent the forward and backward ASE powers, and \( \Delta\nu \) is the spontaneous-emission bandwidth. In this work, the emission bandwidth is measured to be approximately \(0.7\,\mathrm{nm}\), corresponding to \( \Delta\nu \approx 185\,\mathrm{GHz} \). Using the simulated values of \( g_{\text{eff}} \) and \( \alpha_{\text{eff}} \), together with the measured \( \alpha_p \) and \( \alpha_s \), the longitudinal evolution of the optical powers can be numerically evaluated to model the integrated amplifier performance, as shown in Fig.~2a and Fig.~S6.
\\
\\
\changemarker{
\begin{center}
\centering
\includegraphics[width=\linewidth,trim={0cm 0cm 0cm 0cm},clip]{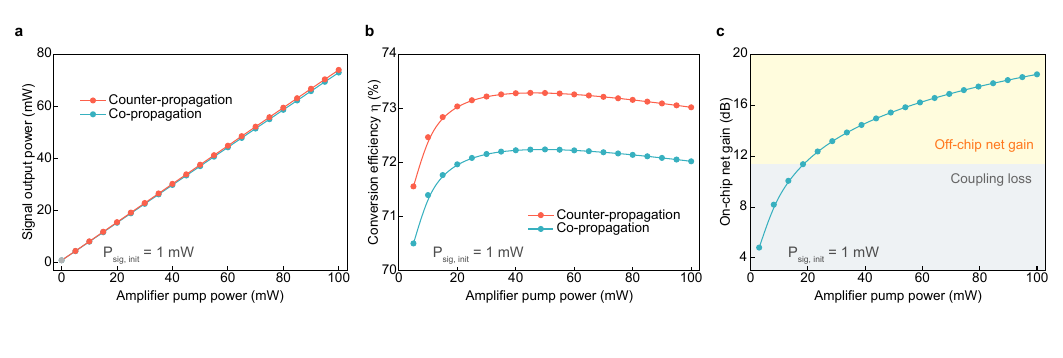}
\begin{flushleft}
\textbf{Fig.~S6. Nd:YAG amplifier. }
\textbf{a}, Simulated signal output power as a function of amplifier pump power for co- and counter-propagating configurations at an input signal power of $P_{sig,init}$ = 1 mW. The two configurations exhibit nearly identical output characteristics.
\textbf{b}, Corresponding conversion efficiency $\eta$ versus amplifier pump power, showing slightly higher efficiency for the counter-propagating configuration due to reduced pump depletion along the propagation direction.
\textbf{c}, Simulated gain for the co-propagating configuration. Shaded areas indicate the regions of off-chip net gain and coupling losses from both facets.
\end{flushleft}
\label{Fig. S6.}
\end{center}
}

For the laser simulation, Eqs.~\ref{eq15} and~\ref{eq16} are omitted, as once lasing begins, stimulated emission dominates over spontaneous emission. In this regime, the out-of-band ASE can be neglected when evaluating the steady-state output power. In the model, the optical pump is assumed to circulate clockwise only, whereas the laser signal propagates in both clockwise and counter-clockwise directions with equal power. The steady-state behavior is determined using an iterative round-trip simulation: in each iteration, the power evolution of both the pump and the signal is computed along the propagation path. The process is repeated until the pump and signal powers converge, reaching self-consistency with the previous iteration. This approach allows evaluation of the steady-state output power for a single output port corresponding to the clockwise signal propagation, as illustrated in Fig.~1b.

\begin{center}
\centering
\includegraphics[width=\linewidth,trim={0cm 0cm 0cm 0cm},clip]{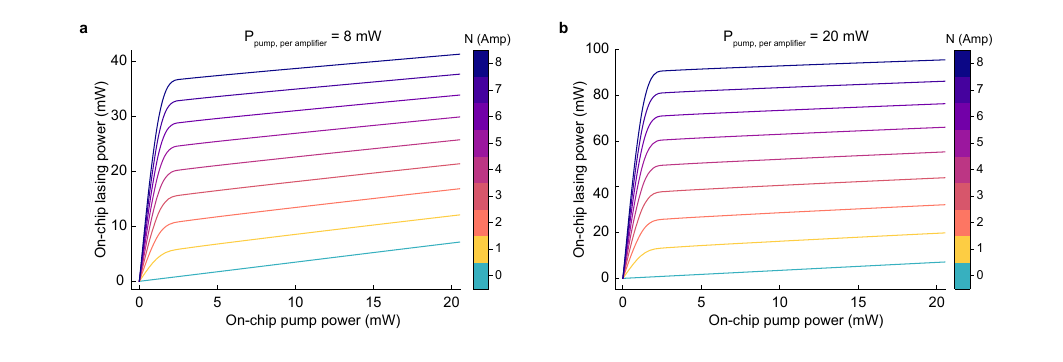}
\begin{flushleft}
\textbf{Fig.~S7. Nd:YAG $\upmu$-MOPA system with additional stages. }
\textbf{a}, Simulated MOPA lasing power versus pump power for varying amplifier stages $N_\mathrm{Amp}$ at $P_\mathrm{pump,~per~amp}=8$~mW. The lasing power exhibits a continuous increase with stage number.
\textbf{b}, Simulations at $P_\mathrm{pump,~per~amp}$ = 20 mW, predicting an output approaching 100 mW with eight amplifier stages.
\end{flushleft}
\label{Fig. S5.}
\end{center}

For the $\upmu$-MOPA system simulations, the microring-generated lasing power and the externally injected pump (both including the additional insertion loss introduced by the on-chip WDM coupler) are used as the input signal and pump for the first amplifier stage. To evaluate the power-scaling capability of the modular $\upmu$-MOPA architecture, we model cascaded amplification for up to eight stages, as shown in Fig.~S7. The conversion efficiency defined as \textit{P}$_\text{lasing}$/\textit{P}$_\text{pump, total}$ for different amplification stages is shown in Fig.~S8.

\begin{center}
\centering
\includegraphics[width=\linewidth,trim={0cm 0cm 0cm 0cm},clip]{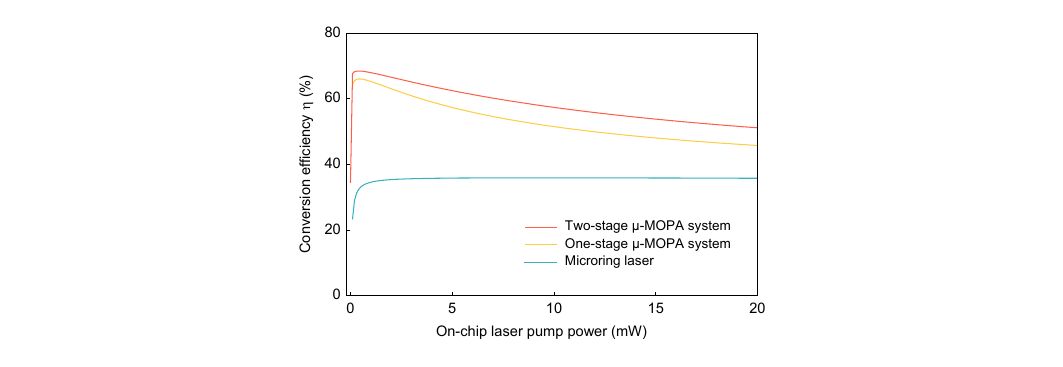}
\begin{flushleft}
\textbf{Fig.~S8. Nd:YAG $\upmu$-MOPA system conversion efficiency. }
Simulated conversion efficiency for the microring laser and $\upmu$-MOPA system under \SI{8}{\milli\watt} pump power per amplifier stage, showing a peak efficiency of 68.4\% for the three-stage $\upmu$-MOPA.
\end{flushleft}
\label{Fig. S6.}
\end{center}

\changemarker{
We further estimated the steady-state temperature rise in the bonded region under both amplification and ring-lasing conditions using finite-element thermal modeling. For the amplifier, we follow the approach of~\cite{yang2024titanium}. In the most conservative estimate, we assume unity quantum efficiency, where each absorbed pump photon is converted into a lasing photon and the residual quantum-defect energy is deposited as heat. The corresponding heat load is given by
\begin{equation}
\frac{dP_{\mathrm{heat}}}{dL} 
= \frac{\hbar \omega_p - \hbar \omega_s}{\hbar \omega_p} \, \alpha_p P_p, \tag{S17}
\end{equation}
where $\alpha_p$ is the pump loss coefficient and $P_p$ is the on-chip pump power. The simulated temperature rise as a function of pump power, up to $500~\mathrm{mW}$, is shown in Fig.~S9a, and the corresponding waveguide cross-sectional temperature distribution at a pump power of $500~\mathrm{mW}$ is shown in Fig.~S9b.}

\changemarker{
For the ring laser, we adopt a complementary estimate based on the generated lasing power. Under the same unity-quantum-efficiency assumption, the absorbed pump power required to generate a lasing power $P_{\mathrm{las}}$ is
\begin{equation}
P_{\mathrm{abs}} = \frac{\omega_p}{\omega_s} \, P_{\mathrm{las}},\tag{S18}
\end{equation}
Therefore, the heat power is
\begin{equation}
P_{\mathrm{heat}} = P_{\mathrm{abs}} - P_{\mathrm{las}} = \left(\frac{\omega_p}{\omega_s} - 1\right) P_{\mathrm{las}}.\tag{S19}
\end{equation}
This heat load was applied to the active ring region in the finite-element simulation. The simulated temperature rise as a function of lasing power, up to 100~mW, is shown in Fig.~S9c, while the corresponding waveguide cross-sectional temperature distribution at a lasing power of 100~mW is shown in Fig.~S9d.}

\changemarker{
For the ring laser, the maximum temperature increase is only $\sim$ 1~K at an on-chip lasing power of 10~mW. Additional simulations show that even for an on-chip lasing power of 100~mW, the waveguide-core temperature increases by only $\sim$ 10~K. For the amplifier, the temperature rise is similarly small. These results are consistent with the absence of thermal rollover in our measurements and support the high conversion efficiency reported here. Such temperature variations are also not expected to substantially affect the gain medium. Previous studies have shown that the quantum efficiency of Nd:YAG crystals at room temperature remains nearly unchanged over temperature changes of this scale~\cite{DEVOR1983NdYAG}.}

\changemarker{
\begin{center}
\centering
\includegraphics[width=\linewidth,trim={0cm 0cm 0cm 0cm},clip]{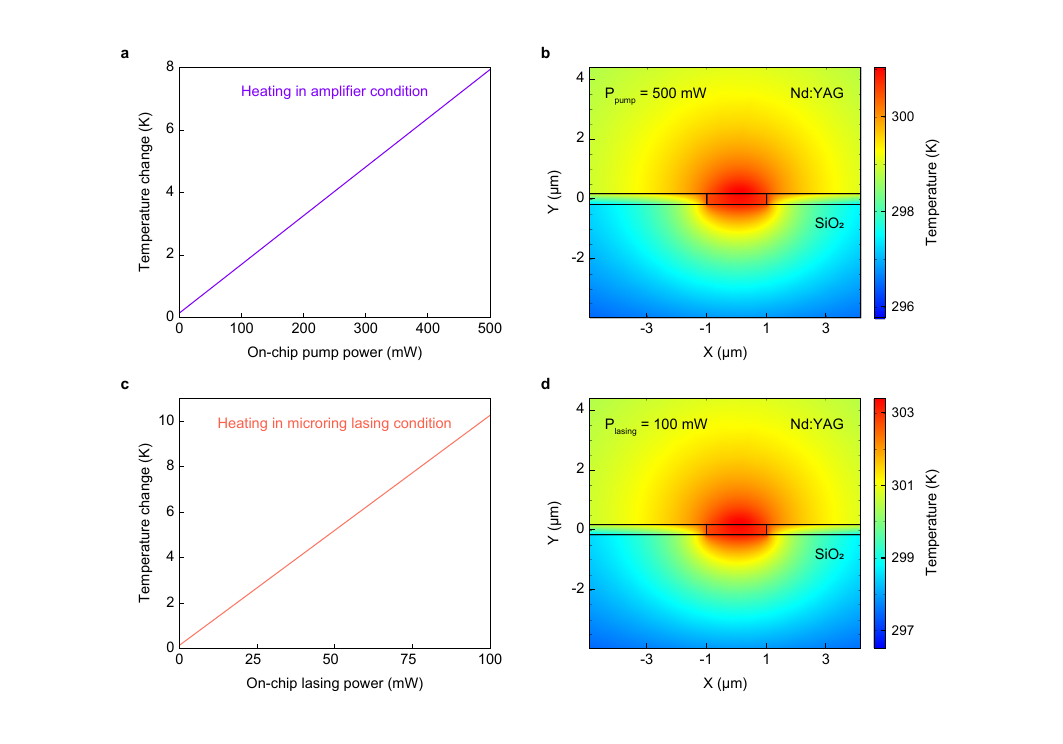}
\begin{flushleft}
\textbf{Fig.~S9. Thermal simulation of the Nd:YAG-bonded Si$_3$N$_4$ waveguide.}
\textbf{a,b}, Simulated temperature rise versus on-chip pump power and corresponding temperature distribution across the waveguide cross-section at 500 mW pump power (amplifier condition).
\textbf{c,d}, Simulated temperature rise versus on-chip lasing power and corresponding temperature distribution across the waveguide cross-section at 100 mW lasing power (microring lasing condition). 
\end{flushleft}
\label{Fig. S6.}
\end{center}
}

\section{Heterodyne spectrum measurement}\label{si:note3}

Heterodyne spectrum measurement is based on coherent detection, where the received optical signal is mixed with a continuous-wave local oscillator (LO), which is a wavelength-swept laser that sweeps across the wavelength window. The interference between the signal and the LO converts the optical frequency information into the radio-frequency (RF) domain, allowing the spectral content of the optical signal to be measured electronically with high resolution.

A schematic of the coherent OSA setup is shown in Extended Fig.2. The input signal $E_s(t)$ and a swept-frequency LO $E_{LO}(t)$ are combined using a 2$\times$2 3-dB fiber coupler. The two output ports are directed to a pair of balanced photodiodes (BPD), and the differential photocurrent is processed through a band-pass electrical filter (BPF), a power monitor (PM), and data acquisition system (DAQ). 

The optical field of the received signal can be expressed following~\cite{HUI2009129}:
\begin{equation}
E_s(t) = A_s(t)\, e^{j(2\pi \nu_s t + \phi_s(t))},
\tag{S20}
\end{equation}
where $A_s(t)$ and $\phi_s(t)$ are the amplitude and phase of the signal, and $\nu_s$ is the mean optical frequency.

The local oscillator is a tunable laser whose optical frequency $\nu_{LO}(t)$ is linearly swept in time, such that
\begin{equation}
E_{LO}(t) = A_{LO}(t) \exp \!\left[ j2\pi \!\int_0^t \! \nu_{LO}(\tau) \, d\tau + j\phi_{LO}(t) \right],
\tag{S21}
\end{equation}
where \(A_{LO}(t)\) and \(\phi_{LO}(t)\) represent the amplitude and phase of the LO, and $\nu_{LO}(t) = \nu_0 + \gamma t$, where $\gamma$ is the frequency sweep rate and $\nu_0$ the initial scanning frequency. The instantaneous phase of the LO can therefore be written as
\begin{equation}
\int_0^t \nu_{LO}(\tau) d\tau = \nu_0 t + \frac{1}{2} \gamma t^2.
\tag{S22} \label{eq22}
\end{equation}

The signal and LO are combined using a 3-dB coupler and detected by a balanced photodetector, which subtracts the two photocurrents to suppress the DC component and laser relative intensity noise. The photocurrent output of the balanced receiver is given by
\begin{equation}
\Delta i(t) = 2 \Re \left\{ R\, A_s(t) A_{LO}(t) 
\, e^{j(2\pi \Delta \nu(t) t + \psi)} \right\},
\tag{S23}
\end{equation}
where $R$ is the detector responsivity, $\Delta \nu(t) = \nu_s - \nu_{LO}(t)$ is the intermediate (beat) frequency, and $\psi = \phi_s(t) - \phi_{LO}(t)$ is the combined phase term.

Substituting Eq.~\ref{eq22}, we obtain a time-varying beat signal whose instantaneous frequency depends quadratically on time:
\begin{equation}
\Delta i(t) = 2 R |A_s(t)| |A_{LO}(t)| 
\cos(\pi \gamma t^2 + \psi).
\tag{S23}
\end{equation}

This is the heterodyne photocurrent, a chirped RF signal that encodes the frequency difference between the signal and the sweeping LO. 

With a band-pass filter included in the electrical receiver, which has a Gaussian temporal response characterized by bandwidth $B$, the photocurrent envelope becomes
\begin{equation}
\Delta i(t) = 2 R |A_s(t)| |A_{LO}(t)| 
\exp\!\left[-\left(\frac{t}{\tau}\right)^2\right]
\cos(\pi \gamma t^2 + \psi),
\tag{S24}
\end{equation}
where $\tau = B / \gamma$ defines the effective temporal resolution corresponding to one optical frequency bin.

As the local oscillator frequency $\nu_{LO}(t)$ is linearly scanned, each instant of time $t$ corresponds to a specific optical frequency $\nu = \nu_0 + \gamma t$. Hence, the time-domain photocurrent $\Delta i(t)$ can be interpreted as a frequency-domain sampling of the signal field $E_s(\nu)$. The envelope of the interference fringes in Eq.~S24 represents the spectral density of the input optical field, while the fast oscillating term $\cos(\pi \gamma t^2 + \psi)$ corresponds to the heterodyne beating process. This scheme converts the optical spectrum into a low-frequency RF beatnote, whose frequency can be measured with sub-MHz precision. Because the resolution is set by electronic sampling rather than grating or slit limitations, the heterodyne meter achieves sub-picometer spectral resolution, far exceeding that of conventional diffraction-based OSAs.


\changemarker{\section{Noise characterization of single-ring and $\upmu$-MOPA lasers}}\label{si:note4}

\changemarker{
To evaluate the spectral coherence and noise performance of the $\upmu$-MOPA, we perform measurements based on heterodyne frequency-noise characterization using an independent reference laser and delayed self-heterodyne measurements using an acousto-optic modulator (AOM) and a long fiber delay (Fig.~S10a). These two techniques probe different aspects of the laser coherence and together provide a comprehensive assessment of the impact of the amplifier stage.
}

\changemarker{
For the heterodyne measurement, the output of the device under test is combined with a narrow-linewidth reference laser (Newport TLB-6721), generating an RF beatnote corresponding to the frequency difference between the two lasers. The detected signal is digitized and processed to extract the phase evolution of the beat note using Welch's method~\cite{welch1967use}. From the phase time series, the single-sideband phase-noise power spectral density $S_{\phi}(f)$ is obtained, which is converted to the frequency-noise PSD $S_f(f)$ via
\begin{equation}
S_f(f) = \frac{f^2}{(2\pi)^2} S_{\phi}(f).\tag{S25}
\end{equation}
The intrinsic linewidth is then inferred from the white frequency-noise floor $h_0$ according to
\begin{equation}
\Delta \nu_{\mathrm{intrinsic}} = \pi h_0.\tag{S26}
\end{equation}
The frequency noise power spectral density $S_f(f)$ was shown in Fig.~S10b. We observe a slight reduction in the white frequency-noise floor after the amplifier stage (from 800 Hz$^2$/Hz to 70 Hz$^2$/Hz). We believe this reduction is attributed to improved signal-to-noise ratio in the heterodyne measurement, rather than a fundamental change in the laser dynamics. 
}

\changemarker{
Independently, we perform delayed self-heterodyne measurements using a 1 km fiber delay line. In this configuration, the laser output is split into two paths, with one arm frequency shifted by the AOM and the other delayed by the fiber. The two paths are recombined on a photodetector to generate an RF beat note. The fiber delay corresponds to a time delay of $\tau \approx 5~\mu\mathrm{s}$, setting a characteristic frequency scale of $1/\tau \sim 200~\mathrm{kHz}$. This measurement is therefore sensitive to phase and frequency noise in the kHz–100 kHz range, where technical noise and ASE-induced fluctuations would be expected to manifest.
}

\changemarker{
The RF beat note is measured on an electrical spectrum analyzer (ESA), and the linewidth is extracted by direct fitting. We obtain an effective beat-note linewidth of approximately 16.0 kHz for the single-ring seed laser. Here, no observable change in the beat-note linewidth is detected after amplification, indicating that the amplifier does not introduce measurable spectral broadening or additional technical noise in this frequency range.
}

\changemarker{
The heterodyne and delayed self-heterodyne measurements probe complementary frequency regimes of the laser noise. The heterodyne frequency-noise PSD provides access to the white-noise floor and thus the intrinsic linewidth, while the delayed self-heterodyne measurement captures technical noise contributions at lower offset frequencies. The absence of any increase in either the white frequency-noise floor or the delayed self-heterodyne linewidth demonstrates that the amplifier preserves both the intrinsic coherence and the practical spectral purity of the seed laser.
}

\changemarker{
Although a travelling-wave amplifier can, in principle, generate broadband ASE across the gain bandwidth, the present $\upmu$-MOPA operates in a strongly seeded regime, where the stimulated signal dominates over spontaneous emission. The combined measurements therefore confirm that ASE does not measurably degrade the coherence of the amplified output.
}
\changemarker{
\begin{center}
\centering
\includegraphics[width=\linewidth,trim={0cm 0cm 0cm 0cm},clip]{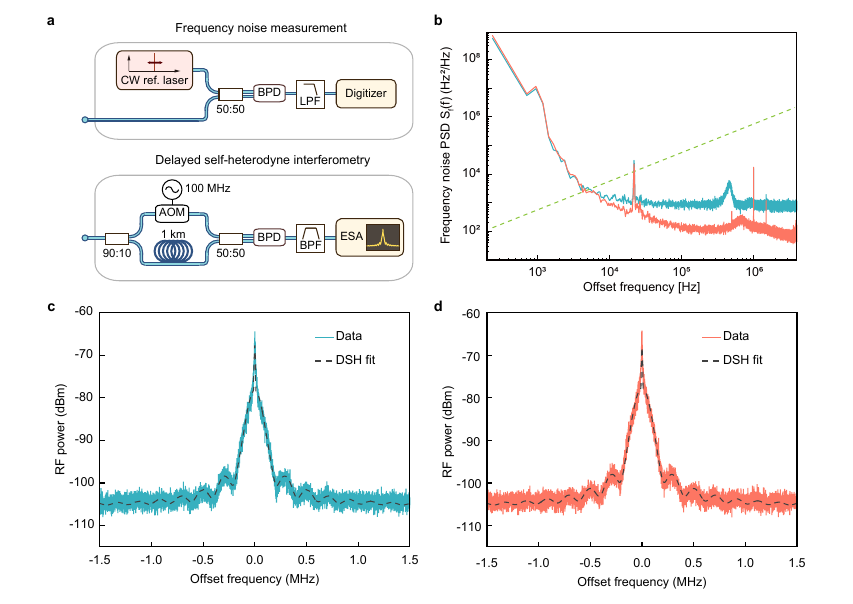}
\begin{flushleft}
\textbf{Fig.~S10. Frequency noise and linewidth characterization of microring laser and $\upmu$-MOPA systems}
\textbf{a}, Schematic of the experimental setup used for frequency noise and linewidth characterization of the microring laser and $\upmu$-MOPA system. The frequency noise is measured via heterodyne detection with a narrow-linewidth reference laser using a BPD, with the signal recorded and analyzed on a digitizer. The linewidth is characterized using delayed self-heterodyne (DSH) interferometry with a 1 km fiber delay.
\textbf{b}, Frequency noise power spectral density $S_f(f)$ of the microring laser (blue) and $\upmu$-MOPA system (orange). The dashed line indicates the $\beta$-separation line.
\textbf{c,d}, Retrieved RF beatnote spectra from delayed self-heterodyne measurements of the (c) microring laser and (d) $\upmu$-MOPA system. Solid lines show the measured RF spectra, and dashed lines indicate the DSH fits. The extracted intrinsic linewidths are 16.0 kHz and 18.9 kHz for the microring laser and $\upmu$-MOPA system, respectively.
\end{flushleft}
\label{Fig.S4}
\end{center}
}

\clearpage
\phantomsection
\section*{Supplementary References}\label{si:references}
\addcontentsline{toc}{section}{Supplementary References}


\begingroup
\makeatletter
\renewcommand{\bibsection}{}%
\makeatother

\endgroup

\clearpage

\end{document}